\documentclass{PoS}

\usepackage{bbm,amsmath} %maths
\usepackage{verbatim,color}

\def\sbar{\overline{s}}
\def\mcC{\mathcal{C}}
\def\mcN{\mathcal{N}}
\def\mcO{\mathcal{O}}

\newcommand{\bp}{\mathbf{p}}
\newcommand{\bq}{\mathbf{q}}
\newcommand{\ubar}{\overline{u}}

\newcommand{\be}{\begin{equation}}
\newcommand{\ee}{\end{equation}}
\newcommand{\bea}{\begin{eqnarray}}
\newcommand{\eea}{\end{eqnarray}}

\newcommand{\bi}{\begin{itemize}}
\newcommand{\ei}{\end{itemize}}

\boldmath
\title{Beyond-the-Standard-Model matrix elements with the gradient flow}

\unboldmath

\ShortTitle{BSM matrix elements with the gradient flow}
\author{
\vspace*{-5mm}
\speaker{A. Shindler}, J. de Vries, T. Luu \\
IAS, IKP, JCHP and JARA-HPC, Forschungszentrum J\"ulich, 52428 J\"ulich, Germany \\
E-mail: \email{a.shindler@fz-juelich.de}, \email{j.de.vries@fz-juelich.de}, \email{t.luu@fz-juelich.de}
}

\abstract{
At the Forschungszentrum J\"ulich (FZJ) we have started a long-term program 
that aims to determine beyond-the-Standard-Model (BSM) matrix elements 
using the gradient flow, and to understand the impact of BSM physics in nucleon and nuclear observables.
Using the gradient flow,
we propose to calculate the QCD component of key
beyond the Standard Model (BSM) matrix elements related to quark and strong $\theta$ CP violation
and the $s\bar{s}$ content within the nucleon.
The former set of matrix elements impacts our understanding of Electric
Dipole Moments (EDMs) of nucleons and nuclei (a key signature of BSM physics),
while the latter contributes to elastic recoil of Dark Matter particles off nucleons and nuclei.
If successful, these results will lay the foundation for extraction of
BSM observables from future low-energy, high-intensity and high-accuracy experimental measurements.
}

\FullConference{The 32nd International Symposium on Lattice Field Theory,\\
23-28 June, 2014\\
Columbia University New York, NY}

\begin{document}
\section{Introduction}
\vspace{-0.3cm}
Sources of nuclear and nucleon EDM within the standard model include 
the CP violation phase in the CKM matrix and the presence of a $\theta$-term in the QCD
Lagrangian. The former source generates a neutron EDM (nEDM) that is 5-6 orders of magnitudes smaller
than the current experimental upper bound. However many extensions of the SM predicts a nEDM
just below the current experimental sensitivity. 
Therefore, with the current experimental accuracy, any measured EDM will be a direct probe 
of unmeasured sources of CP violation. 
Large efforts are underway to improve limits or to find EDMs. Here we want
to mention the experimental proposal for the determination of charged
particles EDMs in the FZJ~\cite{Pretz:2013us}.
A signal in any of the upcoming experiments is either due to the aforementioned $\theta$-term 
or due to BSM physics. The latter can be parametrized by higher-dimensional CP-odd operators.
An important task is then to calculate hadronic and nuclear EDMs in terms of the $\theta$-term and 
these higher-dimensional operators.

\section{EDMs from the $\theta$-term}
\vspace{-0.3cm}

In Euclidean space-time the QCD Lagrangian with a $\theta$ term~\footnote{See~\cite{Vicari:2008jw} for a review.} 
for $N_f$ flavors of fermions $\psi$ reads
\be
{\mathcal L}_\theta= \frac{1}{4} F_{\mu\nu}^a(x)F_{\mu\nu}^a(x)
+ \overline{\psi}\, D \psi
+ \overline{\psi}_L \,M\,\psi_R + \overline{\psi}_R \,M^\dagger\,\psi_L
- i \theta_q q(x),
\end{equation}
where $M$ is the quark mass matrix and $q(x)$ is the topological charge density. 
The imaginary part of the quark mass and the parameter
$\theta_q$ are not independent, i.e. the physical value of $\theta$ is given by
$\theta = \theta_q + {\rm arg}\,{\rm det}\, M$.
This term is CP-violating and induces, for example, an EDM of the nucleon. 
The most stringent constraint on possible CP violation is inferred from measurements of the neutron EDM
(nEDM) $d_N$ and the upper bound $|d_N| < 2.9 \times 10^{-26}\,e\,{\rm cm}\,$ is the 
experimental result of ref.~\cite{Baker:2006ts}.

A computation of the nEDM from QCD with the $\theta$-term combined with the above experimental
bound provides an upper bound in $\theta$.
Assuming $\theta$ is small the nEDM can be written as $d_N \simeq d_N^{(1)} \,\theta \,e\,{\rm fm}$.
The quantity $d_N^{(1)}$ has been calculated in various models 
(for a review, see ref.~\cite{Engelreview}) and approaches based on CP-odd chiral 
Lagrangians.
Lattice QCD data (see ref.~\cite{Shintani:2008nt} 
for older lattice QCD results) have been analyzed 
in the chiral approach providing the most recent values of $d_N^{(1)}$ so far~\cite{Savage,Guo:2012vf,Guo2}. 
Taking the lower bound of all these approaches, the experimental upper bound
for $d_N$ leads to $|\theta| \lesssim {\rm O}(10^{-10})$.
More accurate lattice calculations of $d_N^{(1)}$ (and BSM CP-odd matrix elements) 
are necessary to set better limits and to identify the fundamental CP-odd source 
from future nonzero EDM measurements~\cite{Dekens}. 

\subsection{Lattice details}
%\vspace{-0.2cm}
The relevant matrix element to obtain the dipole moment of nucleons in a  
$\theta$ vacuum is
\bea
\phantom{a}_\theta\langle \bp_2,s_2| J_{\rm em}^\mu | \bp_1,s_1 \rangle_\theta = 
\bar{u}_{N}^\theta(\bp_2,s_2) \Gamma^\mu(q^2) u_N^\theta(\bp_1,s_1), \label{eleform}
\eea
where $\Gamma^\mu(q^2)$, with $q=p_2-p_1$, is a linear combination of all
possible form factors consistent with the symmetries of QCD with a $\theta$-term, i.e.
gauge, Lorentz, and CPT symmetry
\be
\Gamma^\mu(q^2) = F_1(q^2) \gamma^\mu  + 
\frac{1}{2M_N} F_2(q^2) i\sigma^{\mu\nu} q_\nu 
 + F_A(q^2) (\gamma^\mu \gamma^5 q^2 - 2 M_N \gamma^5 q^\mu) +
\frac{1}{2 M_N} F_3(q^2) \sigma^{\mu\nu}\gamma_5 q_\nu\,.
\ee
The electric dipole moment which vanishes when $\theta\to
0$ is given by $d_N = F_3(0)/2M_N$.
The form factors can be extracted, in lattice calculations, from the
large Euclidean time behavior of the following three-point function
\be
\langle \mcN(t_2,\bp_2) J_{\rm em}^\mu(t,\bq) 
\mcN^\dagger(t_1,\bp_1) \rangle_\theta,
\ee
where $\mcN$ is the nucleon interpolating operator 
and $J_{\rm em}^\mu$ is an electromagnetic current insertion.

Numerical Monte Carlo methods cannot be directly applied
to the action with $\theta\ne 0$.
The small value of $\theta$ inferred from the experimental bounds
allows us to expand a generic expectation value of an operator $\mathcal{O}$
in a $\theta$-vacuum
\be
\langle  \mathcal{O} \rangle_\theta \simeq
\langle  \mathcal{O} \rangle_{\theta=0} + i \theta \langle \mathcal{O} Q \rangle_{\theta=0} + \mathcal{O}(\theta^2)
\label{eq:O_theta}
\ee
where $Q$ is the topological charge. 
Retaining only the linear term in $\theta$ one can obtain the desired result for $d_N^{(1)}$.
Another possible approach to determine the nucleon EDM 
at finite $\theta$ is to use reweighting techniques with
the complex weight factor $e^{i\theta Q}$. 

We propose to compute directly the matrix element in eq.~\eqref{eq:O_theta} using
the gradient flow~\cite{Luscher:2010iy} to define the topological charge. 
This allows us to perform a safe continuum limit to all correlators that
contain the topological charge without encountering difficult renormalization
patterns. It also allows to use any fermion lattice action
without being constrained to the use of a Ginsparg-Wilson fermions.

As a preparatory work we compute the topological susceptibility
at non-vanishing flow-time. This provides us with a better understanding of the 
behavior of these local operators as a function of the flow-time.

\subsection{Topological susceptibility}
%\vspace{-0.2cm}
As a first check of the setup of our 
calculation we determine the topological susceptibility.
For the topological charge 
\be
Q(t) = \int d^4 x~q(x,t)\,, \qquad 
q(x,t) = - \frac{1}{32\pi^2} \epsilon_{\mu\nu\rho\sigma} {\rm tr}\left\{G_{\mu\nu}(x,t)G_{\rho\sigma}(x,t)\right\}\,,
\ee
we use the gauge field definition, where the topological charge density $q(x,t)$
has been computed with gauge fields evolved using the gradient flow~\cite{Luscher:2010iy}.
Details and definitions about the gradient flow of gauge fields can be found in~\cite{Luscher:2010iy}.

\begin{figure}[tb]
\begin{center}
\vspace{-0.8cm}\includegraphics[width=8cm]{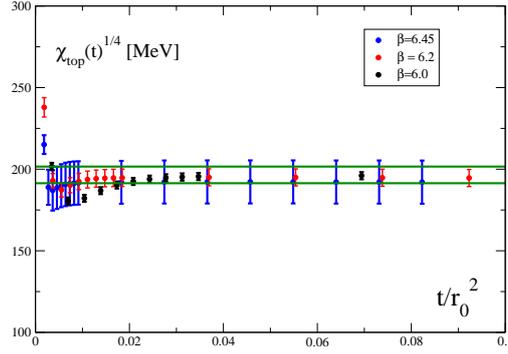}
\end{center}
\vspace{-1.4cm}
\caption{Flow-time dependence of the topological susceptibility at 3 different
lattice spacings. The green band is the value of ref.~\cite{Luscher:2010iy}.}
\label{fig:topo_susc}
\end{figure}

In the continuum limit for every $t>0$ we expect the topological susceptibility to be independent
of the flow-time $t$. We can perform a safe continuum limit at a given fixed value
of the flow-time $t$ and the final result should be independent
of $t$. In fig.~\ref{fig:topo_susc} we show the flow-time dependence
in units of $r_0$ of the topological susceptibility. Excluding a very small region
close to the $t=0$ boundary where cutoff effects dominate,
the topological susceptibility is flow-time independent
and suffers, within the current statistical precision, with small discretization
uncertainties. For comparison the green band is the value given in ref.~\cite{Luscher:2010iy}
after performing the continuum limit.

\subsection{Mixing in the CP-broken vacuum}
%\vspace{-0.2cm}
In a $\theta$-vacuum CP is not a symmetry and as a consequence
positive parity states \begin{comment}
as the neutron can
\end{comment}
 mix with negative parity states,
causing $\theta=0$ eigenstates to mix in a CP-broken vacuum.
In a lattice calculation one always deals with correlation functions
and not directly with matrix elements of given definite states, thus 
one must take into account this mixing
when extracting matrix elements between definite states.
In particular, when extracting form factors from correlation functions, 
one must account for the unphysical mixing of the electric and magnetic dipole moments.

The effect of such mixing can already be studied by considering
nucleons' two-point functions in a $\theta$-vacuum.
In the absence of CP invariance nucleons
can have an additional Dirac structure proportional to $i\gamma_5$
due to the phase factor ${\rm exp}i \alpha_N \gamma_5$ that results between the coupling of the 
physical nucleon state $|N\rangle$ and the corresponding interpolation operator.
In terms of the completeness relations
\be
\sum_s u_N^\theta(\bp,s) \ubar_N^\theta(\bp,s) = E_\theta(\bp)\gamma_0 - i \gamma_k p_k + M_N {\rm e}^{2 i \alpha_N(\theta)\gamma_5}\,.
\ee
The phase factor $\alpha_N$ can be determined  by a study of 
the nucleon two-point functions
in a $\theta$-vacuum. Assuming $\theta$ is small, we can expand
\be
G_{NN}^\theta(x_0) = a^3 \sum_{\underline{x}} \left\langle \mcN(\underline{x},x_0)\mcN^\dagger(0)\right\rangle_\theta 
= G_{NN} + i \theta G_{NN}^Q + {\rm O}(\theta^2)\,.
\ee
The first term in the small $\theta$ expansion is the usual nucleon two-point function. Projecting
into positive parity state the leading term in the spectral decomposition is
\be
{\rm tr}\left[P_+G_{NN}\right] = |Z_N|^2 {\rm e}^{-M_N x_0} + \cdots
\label{eq:G_NN0}
\ee
where $M_N$ is the nucleon mass and $Z_N$ is the coupling between the interpolating operator
and the nucleon state in the $\theta=0$ vacuum.
The first correction proportional to $\theta$ once projected with $\gamma_5$ has 
a spectral decomposition with leading term given by~\cite{Shintani:2005xg}
\be
{\rm tr}\left[P_+ \gamma_5 G_{NN}^Q\right] = |Z_N|^2 \alpha_N^{(1)} {\rm e}^{-M_N x_0} + \cdots
\label{eq:G_NN1}
\ee
The exponential decay is still given by the nucleon mass
and the amplitude gets a multiplicative correction factor $\alpha_N^{(1)}$, where
$\alpha_N(\theta) \simeq \theta \alpha_N^{(1)} + \cdots\,$.

The fact that both correlation functions have the same leading exponential decay 
allows us to check the correctness of the sampling of all topological sectors and the space-time overlap 
between the nucleon propagator and the topological charge.
It also allows to extract $\alpha_N^{(1)}$ from the ratio of the two correlators.
\be
\alpha_N^{(1)} = \frac{{\rm tr}\left[P_+ \gamma_5 G_{NN}^Q\right]}{{\rm tr}\left[P_+G_{NN}\right]} + \cdots
\label{eq:ratio}
\ee

\begin{figure}[tb]
\vspace{-0.8cm}\hspace{-1.cm}
\includegraphics[width=8.5cm]{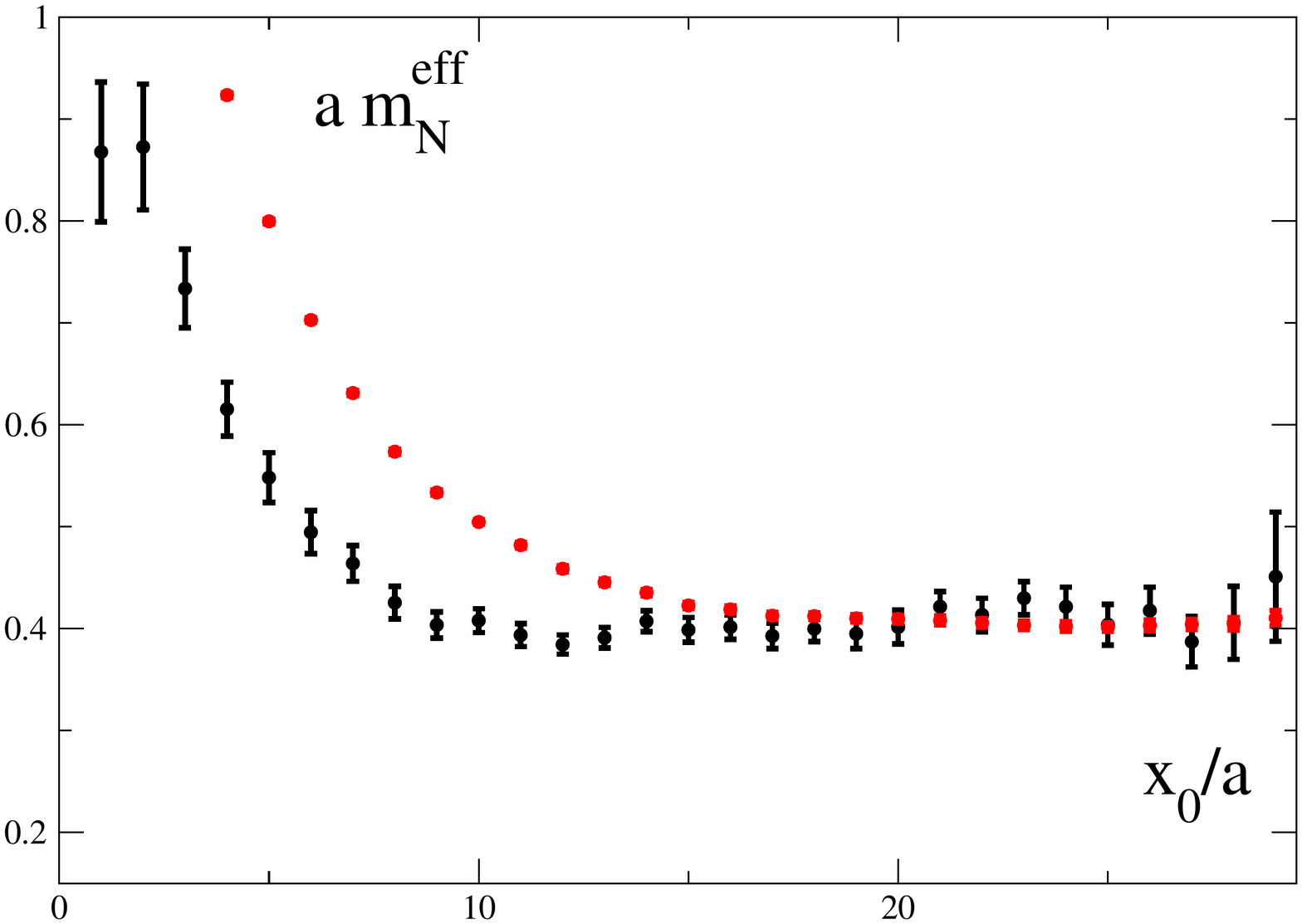}
\includegraphics[width=8.5cm]{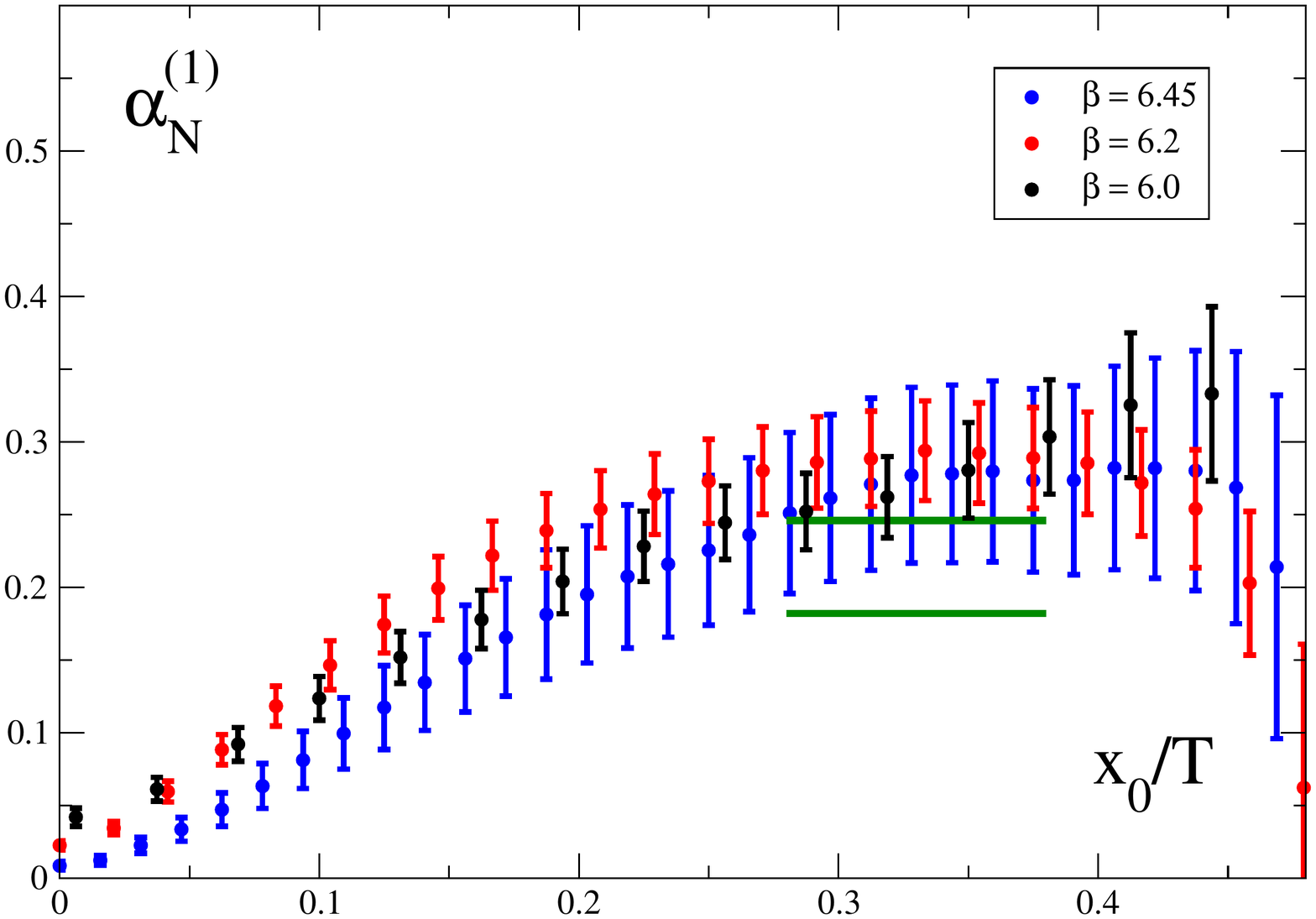}
\vspace{-1.4cm}
\caption{Left plot: nucleon effective mass coming from the correlator in eq.~\protect\eqref{eq:G_NN0}
(red points) and from the correlator in eq.~\protect\eqref{eq:G_NN1} (black points). 
Right plot: Euclidean time dependence of the ratio~\protect\eqref{eq:ratio} at two different
lattice spacings. The blue band is the value of $\alpha_N^{(1)}$ obtained in \cite{Berruto:2005hg} using a different
fermion and gauge lattice action.}
\label{fig:nucleon}
\end{figure}

In the left plot of fig.~\ref{fig:nucleon} we compare the nucleon effective masses at $\beta=6.45$
extracted from the correlator in eq.~\eqref{eq:G_NN0} (red points) and the correlator~\eqref{eq:G_NN1}.
We observe that at large Euclidean time the effective masses are well compatible
within our statistical accuracy.
This agreement shows that we properly sample 
all topological sectors and confirms the expectations from eqs.~(\ref{eq:G_NN0}, \ref{eq:G_NN1}).
It is interesting to observe that the correlator~\eqref{eq:G_NN1} is noisier than the
correlator~\eqref{eq:G_NN0} but is less contaminated by excited states.

We can now compute the ratio~\eqref{eq:ratio} that it is shown in the right plot of fig.~\ref{fig:nucleon}
at 2 lattice spacings, $\beta=6.45$ (black points) and $\beta=6.0$ (red points). At $\beta=6.45$
we can safely extract the value of the parameter $\alpha_N^{(1)}$ and even at $\beta=6.0$ 
we can observe a reasonable plateau. We notice that $\alpha_N^{(1)}$ does not seem 
to suffer from big cutoff effects within our current statistical uncertainties and it agrees reasonably well
with the value of $\alpha_N^{(1)}$ obtained in~\cite{Berruto:2005hg} using a different
fermion and gauge lattice action.
There is little doubt that we can safely perform a continuum limit both
for the topological susceptibility and the phase factor $\alpha_N^{(1)}$.

\section{Dark matter}
\vspace{-0.3cm}
A large class of models~\cite{Bertone:2004pz} predict 
weakly interacting massive particle (WIMP) as dark matter (DM) candidate.
Ongoing direct-detection experiments provide severe constraints on the parameters space of such models.

The WIMP couples, via the exchange of a Higgs boson, to the various quark scalar density operators 
taken between nucleon states.
This provides a scenario for the detection of a WIMP type of DM particles.
The cross section for spin independent elastic WIMP--nucleon ($\chi N$) scattering 
at zero momentum transfer~\cite{Ellis:2008hf} is proportional to
\be
\Big\lvert\sum_{f} G_f f_{T_f} \Big\rvert^2  \qquad \text{with} \qquad f_{T_f} 
= \frac{m_{f}}{M_N} \left\langle N | \bar{q}_f{q_f} |N \right\rangle \, .
\label{eq:crosssection} 
\ee
The dependence on the parameters of the particular BSM theory used for the calculation of
the cross section is contained in the function $G_f$.
The dimensionless and renormalization-group 
invariant (RGI) coupling $f_{T_f}$ depends on the mass $m_{f}$ of the quark 
of flavor $f$ and the nucleon mass, $M_N$.  
As evident from Eq.~(\ref{eq:crosssection}), the cross section
depends quadratically on $f_{T_f}$, and is therefore very sensitive 
to the size of the scalar content contributions of different flavors. 

In principle lattice QCD can provide a determination 
of these nucleon matrix elements from first principles. 
Here we propose a new method, based on the gradient flow for fermions~\cite{Luscher:2013cpa},
that will reduce some of the uncertainties of previous calculations.

\subsection{A new method}
%\vspace{-0.2cm}
The method is based on the possibility to relate the matrix element of 
a scalar density 
at non-vanishing flow-time between nucleon states
with the the one at vanishing flow-time. 
This can be done considering the small flow-time expansion
or finding the proper Ward identities (WI) as for the case
of the chiral condensate~\cite{Luscher:2013cpa,Shindler:2013bia}.
As an example we briefly discuss how the method works
in the case of the strange content of the nucleon, but the method
is applicable to any quark content of the nucleon.
The small flow-time expansion of the strange scalar density
$\mcO_s(t,x) = \sbar(t,x)s(t,x)$
in the continuum theory reads
\be
\mcO_s(t,x) = c_0(t)m_s + c_1(t)m_s\left(m_u^2+m_d^2+m_s^2\right) + c_2(t)m_s^3 + c_3(t)\mcO_s(0,x) + O(t)
\ee
where $c_0 \sim 1/t$ and $c_{1,2,3} \sim \log t$ and $m_u,m_d,m_s$ are the 
quark masses.
If one considers the subtracted matrix element
\be
\mcC^{\rm sub}(t,x) = \left\langle\mcN \mcO_s(t,x) \mcN^\dagger\right\rangle - \left\langle\mcO_s(t,x) \right\rangle \left\langle\mcN \mcN^\dagger \right\rangle
\ee
the small flow-time expansion contains only the term proportional to $c_3$
\be
\mcC^{\rm sub}(t,x) = c_3(t)\mcC^{\rm sub}(0,x) +O(t)\,.
\ee
At finite flow-time no mixing is present.
To compute the physical matrix element at $t=0$ one needs to compute $c_3(t)$.
Chiral symmetry implies that the leading coefficient of the small flow-time
expansion of the pseudoscalar density also is $c_3(t)$, thus 
using the spectral decomposition of pseudoscalar two-point functions
with operators at vanishing and non-vanishing flow-time one obtains
$c_3(t) = \frac{G_{\pi,t}}{G_\pi} + O(t)\,$,
i.e. a non-perturbative determination of the $c_3(t)$ coefficient.
This implies that to compute the matrix element related to the strange content of the nucleon 
one needs to compute
\be
\mcC^{\rm sub}(0,x) = \frac{G_{\pi}}{G_{\pi,t}} \cdot \left[
\left\langle\mcN \mcO_s(t,x) \mcN^\dagger\right\rangle - \left\langle \mcO_s(t,x) \right\rangle \left\langle\mcN \mcN^\dagger \right\rangle\right] + O(t)\,.
\label{eq:sbars}
\ee
We observe that the renormalization factor of $\mcO_s(t,x)$ simplifies with the one of $G_{\pi,t}$, thus the
only renormalization factor needed is the one for $G_\pi$, i.e. $Z_P$.
The matrix element~\eqref{eq:sbars} should be computed in a window of $t$-values,
not too small to avoid big discretization errors and not too large to avoid contributions
from higher dimensional operators. The existence of the window can be systematically
checked because the continuum limit can be safely performed at fixed physical value
of $t$ and because once the continuum limit is performed, the matrix element~\eqref{eq:sbars} should be 
independent of $t$ if the small flow-time expansion does not receive contributions
from higher dimensional operators.
The advantages of this method are that at all steps of the calculation
the theory is unitary. There is no mixing for the operators defined
at positive flow-time at all stages of the calculation.
Using twisted mass fermions~\cite{Frezzotti:2000nk} at maximal twist the matrix elements in~\eqref{eq:sbars}
defined at positive flow-time are automatic O($a$) improved~\cite{Frezzotti:2003ni,Shindler:2013bia}.

For future $N_f>0$ calculations, of both the EDM and the strange content of the nucleon,
we are planning to use gauge configurations available on the lattice
data grid.

\subsection*{Acknowledgments}
%\vspace{-0.2cm}
A.S. acknowledges enjoyable discussions with M. L\"uscher, M. Papinutto, A. Patella
and N. Tantalo and many important email exchanges with R. Edwards and B. Joo.
Work is supported in part by DFG and NSFC (CRC 110) (JdV). 
The authors gratefully acknowledge the computing time granted by the JARA-HPC 
Vergabegremium and provided on the JARA-HPC Partition part of the supercomputer 
JUQUEEN at Forschungszentrum J\"ulich.

\bibliographystyle{h-elsevier}
\bibliography{lat14}

\end{document}